\documentclass[twocolumn,showpacs,preprintnumbers,amsmath,amssymb]{revtex4}
 
\usepackage{graphicx}
\usepackage{dcolumn}
\usepackage{bm}
 
\newcommand{\mean}[1]{\langle{#1}\rangle}

\newcommand{\dgg}{^{\dagger}}
\newcommand{\Tr}{{\rm Tr}\hspace{0.07cm}}

\newcommand{\im}{{\rm i}}

\newcommand{\half}{\frac{1}{2}}

\newcommand{\condbar}{\hspace{0.1em}|\hspace{0.1em}}

\begin{document}
 
\preprint{APS/123-QED}
 
\title{Relation between fundamental estimation limit and stability 
in linear quantum systems with imperfect measurement
}
 
\author{Naoki Yamamoto}
 \email{naoki.yamamoto@anu.edu.au}
\affiliation{%
Department of Engineering, Australian National University, 
ACT 0200, Australia
}

\author{Shinji Hara}
 \email{Shinji_Hara@ipc.i.u-tokyo.ac.jp}
\affiliation{%
Department of Information Physics and Computing, University of Tokyo,
Hongo 7-3-1, Bunkyo-ku, Tokyo 113-0033, Japan
}

\date{\today}

\begin{abstract}

From the noncommutative nature of quantum mechanics, estimation of canonical 
observables $\hat{q}$ and $\hat{p}$ is essentially restricted in its 
performance by the Heisenberg uncertainty relation, 
$\mean{\Delta \hat{q}^2}\mean{\Delta \hat{p}^2}\geq \hbar^2/4$. 
This fundamental lower-bound may become bigger when taking the 
structure and quality of a specific measurement apparatus into account. 
In this paper, we consider a particle subjected to a linear dynamics that 
is continuously monitored with efficiency $\eta\in(0,1]$. 
It is then clarified that the above Heisenberg uncertainty relation is 
replaced by $\mean{\Delta \hat{q}^2}\mean{\Delta \hat{p}^2}\geq \hbar^2/4\eta$ 
if the monitored system is unstable, while there exists a stable quantum 
system for which the Heisenberg limit is reached.

\end{abstract}

\maketitle


In quantum mechanics, any noncommutative observables must possess a 
fundamental uncertainty due to the absence of their joint probability 
distribution. 
For example, if we estimate the position and momentum operators of a single 
particle, $\hat{q}$ and $\hat{p}$, the estimation errors $\Delta\hat{q}$ and 
$\Delta\hat{p}$ satisfy the Heisenberg uncertainty relation 
$\mean{\Delta \hat{q}^2}\mean{\Delta \hat{p}^2}\geq \hbar^2/4$. 
Several type of such uncertainty bounds have been found in quite general 
formulation that even includes effects of measurement \cite{ozawa1,ozawa2}. 
It is clearly significant to perform further detailed investigation on 
fundamental estimation limit taking the structure, properties, and quality 
of a specific estimator into account.

The {\it quantum filter} \cite{belavkin1,belavkin2,belavkin3,luc1,luc2} 
is a particularly important estimator, because of its potential application 
to quantum feedback control 
\cite{thomsen,ahn,geremia,stockton,ramon,james,mirrahimi,yamamoto1}. 
More specifically, for a continuously monitored system, the quantum filter 
generates an optimal estimate of a system observable, which can be fed back 
to control the system. 
The estimator is recursively computed using the {\it Belavkin filtering 
equation}; this completely reflects the structure of the monitored system. 
Hence, within the framework of quantum filtering, the estimation limit is 
determined by dynamical properties of the system, e.g., the stability.

In this paper, we particularly focus on a single one-dimensional particle 
that has a quadratic potential and a linear interaction with a vacuum 
electromagnetic field, the latter of which is continuously measured by 
a homodyne detector \cite{yanagisawa,doherty1,doherty2,wiseman,gough,
simon,yamamoto2,wilson}. 
For this system, the filtering equation is reduced to the famous 
{\it Kalman filter}, and eventually the estimation error can be evaluated 
explicitly. 
The goal of this paper is to show that, irrespective of parameters of the 
system, there exists a fundamental estimation limit determined by the 
dynamical stability properties of the system. 
In particular, we show that a new estimation limit on $\hat{q}$ and 
$\hat{p}$ appears if the system is unstable, while there exists a stable 
quantum system for which the Heisenberg limit is reached.

We use the following notation: 
for a matrix $A=(a_{ij})$, the symbols $A^{{\mathsf T}}$, $A\dgg$, and 
$A^*$ represent its transpose, conjugate transpose, and elementwise complex 
conjugate of $A$, i.e., 
$A^{{\mathsf T}}=(a_{ji})$, $A\dgg=(a_{ji}^*)$, 
and $A^*=(a_{ij}^*)=(A\dgg)^{{\mathsf T}}$, respectively; 
these rules are applied to any rectangular matrix including column and 
row vectors. 
${\rm Re}(A)$ and ${\rm Im}(A)$ denote the real and imaginary part of $A$, 
respectively, i.e., $({\rm Re}(A))_{ij}=(a_{ij}+a_{ij}^*)/2$ and 
$({\rm Im}(A))_{ij}=(a_{ij}-a_{ij}^*)/2\im$.

We first review the quantum filtering theory with the focus on a particle 
interacting with a field. 
The interaction is given by a unitary operator subjected to the following 
{\it Hudson-Parthasarathy equation} \cite{hudson}: 
\begin{equation}
\label{HP-eq}
   \hbar d\hat{U}_t
      =\Big[ \big(-\im \hat{H}-\half \hat{c}\dgg \hat{c}\big)dt
                        +\hat{c}d\hat{B}_t\dgg
                        -\hat{c}\dgg d\hat{B}_t \Big]\hat{U}_t,~
   \hat{U}_0=\hat{I}, 
\end{equation}
where $\hat{c}=c_1\hat{q}+c_2\hat{p}$. 
The constants $c_1,~c_2\in{\mathbb C}$ are determined according to the 
system-field interaction. 
The quantum Wiener process $\hat{B}_t$, which is a field operator, satisfies 
the following quantum Ito rule: 
\[
   d\hat{B}_t d\hat{B}_t=0,~d\hat{B}_t\dgg d\hat{B}_t=0,~
   d\hat{B}_t d\hat{B}_t\dgg=\hbar dt,~d\hat{B}_t\dgg d\hat{B}_t\dgg=0. 
\]
In addition to the interaction, the particle is trapped in a quadratic 
harmonic potential of the form
\[
    \hat{H}=\half\hat{x}^{{\mathsf T}}G\hat{x}
           =\half(g_{11}\hat{q}^2+g_{12}\hat{q}\hat{p}
                 +g_{12}\hat{p}\hat{q}+g_{22}\hat{p}^2), 
\]
where $\hat{x}=(\hat{q},\hat{p})^{{\mathsf T}}$, and $G=(g_{ij})$ is a 
$2\times 2$ real symmetric matrix. 
In the Heisenberg picture, the time-evolved position and momentum operators 
$\hat{q}_t=\hat{U}_t\dgg\hat{q}\hat{U}_t$ and 
$\hat{p}_t=\hat{U}_t\dgg\hat{p}\hat{U}_t$ satisfy the following quantum 
stochastic differential equation: 
\begin{equation}
\label{linear-qsde}
    d\hat{x}_t=A\hat{x}_tdt
           +\im\Sigma[Cd\hat{B}_t\dgg-C^* d\hat{B}_t], 
\end{equation}
where $\hat{x}_t=(\hat{q}_t,\hat{p}_t)^{{\mathsf T}}$. 
Here, we have defined 
\[
     A:=\Sigma\big[G+{\rm Im}(C^*C^{{\mathsf T}})\big],~
     C:=
     \left( \begin{array}{c}
         c_1 \\
         c_2 \\
     \end{array} \right),~
     \Sigma:=
     \left( \begin{array}{cc}
         0 & 1 \\
         -1 & 0 \\
     \end{array} \right). 
\]
Next, we consider to measure a field observable after the interaction. 
In the homodyne detection scheme, the observable to be measured is given 
by 
\[
    Y'_t=\hat{U}_t\dgg({\rm e}^{-\im\phi}\hat{B}_t
            +{\rm e}^{\im\phi}\hat{B}_t\dgg)\hat{U}_t
               +\kappa(\hat{B}'_t+\hat{B}_t'\mbox{}\dgg), 
\]
where $\hat{B}_t'$ is a noise uncorrelated from $\hat{B}_t$, and $\kappa\geq0$ 
represents the strength of $\hat{B}_t'$. 
Also, $\phi\in[0,2\pi)$ denotes a phase-shift parameter that should be 
optimized. 
Redefining the normalized output $Y_t$ satisfying $dY_t^2=dt$, we have 
\begin{eqnarray}
& & \hspace*{-1em}
\label{linear-output}
    dY_t=2\sqrt{\eta}C_{{\rm r}}^{{\mathsf T}}\hat{x}_tdt
              +\sqrt{\eta}({\rm e}^{-\im\phi}d\hat{B}_t
              +{\rm e}^{\im\phi}d\hat{B}_t\dgg)
\nonumber \\ & & \hspace*{8em}
         \mbox{} +\sqrt{1-\eta}(d\hat{B}'_t+d\hat{B}_t'\mbox{}\dgg), 
\end{eqnarray}
where $C_{{\rm r}}:={\rm Re}({\rm e}^{-\im\phi}C)$ and 
$\eta:=(1+\kappa^2)^{-1}\in(0,1]$. 
Remarkably, $Y_t$ satisfies the {\it self-nondemolition} property 
$[Y_s,Y_t]=0,~\forall s,t$ for a fixed $\phi$, which indicates that the 
observation ${\cal Y}_t={\rm vN}\{Y_s~|~0\leq s\leq t\}$ constructs a 
classical stochastic process. 
Furthermore, $Y_t$ satisfies the {\it nondemolition} condition 
$[Y_s, \hat{q}_t]=0,~[Y_s, \hat{p}_t]=0,~\forall s\leq t$ for a fixed $\phi$. 
These two properties allow us to define the {\it quantum conditional 
expectations} $\pi_t(\hat{q})={\mathbb P}(\hat{q}_t\condbar{\cal Y}_t)$ and 
$\pi_t(\hat{p})={\mathbb P}(\hat{p}_t\condbar{\cal Y}_t)$, which are the best 
estimates of $\hat{q}_t$ and $\hat{p}_t$ in the sense of the least mean 
square error. 
Following the quantum filtering theory, we obtain a recursive equation to 
calculate $\pi_t(\hat{q})$ and $\pi_t(\hat{p})$: 
\begin{eqnarray}
& & \hspace*{-1em}
\label{linear-filter}
    d\pi_t(\hat{x})=A\pi_t(\hat{x}) dt
\nonumber \\ & & \hspace*{0em}
    \mbox{}+\sqrt{\eta}\Big[\frac{2}{\hbar}V_t C_{{\rm r}}
        +\Sigma^{{\mathsf T}}C_{{\rm i}}\Big]
          \big[dY_t-2\sqrt{\eta}C_{{\rm r}}^{{\mathsf T}}\pi_t(\hat{x}) dt\big], 
\end{eqnarray}
where $C_{{\rm i}}:={\rm Im}({\rm e}^{-\im\phi}C)$ and 
$\pi_t(\hat{x}):=(\pi_t(\hat{q}),\pi_t(\hat{p}))^{{\mathsf T}}$. 
Here, $V_t$ is the symmetrized covariance matrix given by 
\begin{eqnarray}
& & \hspace*{-1em}
\label{covariance}
    V_t:={\mathbb P}(\hspace{0.05cm}\hat{P}_t
              \hspace{0.05cm}|\hspace{0.05cm}{\cal Y}_t\hspace{0.05cm})
\nonumber \\ & & \hspace*{-1em}
    \hat{P}_t:=\left( \begin{array}{cc}
     \Delta\hat{q}_t^2 
      & \half(\Delta\hat{q}_t\Delta\hat{p}_t
             +\Delta\hat{p}_t\Delta\hat{q}_t) \\
     \half(\Delta\hat{q}_t\Delta\hat{p}_t+\Delta\hat{p}_t\Delta\hat{q}_t) 
      & \Delta\hat{p}_t^2 
        \end{array} \right), 
\nonumber \\ & & \hspace*{-1em}
    \mbox{}
\end{eqnarray}
where $\Delta\hat{q}_t:=\hat{q}_t-\pi_t(\hat{q})$ and 
$\Delta\hat{p}_t:=\hat{p}_t-\pi_t(\hat{p})$ are the estimation errors. 
$V_t$ satisfies the following Riccati differential equation: 
\begin{equation}
\label{riccati}
    \dot{V}_t=A'V_t+V_t A'\mbox{}^{{\mathsf T}}+D
              -\frac{4\eta}{\hbar}V_tC_{{\rm r}}C_{{\rm r}}^{{\mathsf T}}V_t, 
\end{equation}
where 
\begin{eqnarray}
& & \hspace*{-1em}
    A':=\Sigma\big[G+C_{{\rm r}}C_{{\rm i}}^{{\mathsf T}}+(2\eta-1)C_{{\rm i}}C_{{\rm r}}^{{\mathsf T}}\big], 
\nonumber \\ & & \hspace*{-1em}
    D:=\hbar\Sigma^{{\mathsf T}}
         \big[C_{{\rm r}}C_{{\rm r}}^{{\mathsf T}}+(1-\eta)C_{{\rm i}}C_{{\rm i}}^{{\mathsf T}}\big]\Sigma. 
\nonumber
\end{eqnarray}
As Eq. (\ref{riccati}) is deterministic, the quantum conditional expectation 
$V_t={\mathbb P}(\hspace{0.05cm}\hat{P}_t
\hspace{0.05cm}|\hspace{0.05cm}{\cal Y}_t\hspace{0.05cm})$ is replaced by 
the simple expectation 
$V_t=\mean{\hat{P}_t}:=\Tr[(\rho\otimes\Phi)\hat{P}_t]$, where $\rho$ is 
a system state and $\Phi$ is the field vacuum state. 
The set of equations (\ref{linear-filter}) and (\ref{riccati}) called 
the {\it quantum Kalman filter} computes the best estimate of $\hat{q}_t$ 
and $\hat{p}_t$ recursively.

We here provide an important fact: 
Unlike the classical case where the error covariance matrix is simply a 
nonnegative matrix, the canonical commutation relation 
$[\hat{q},\hat{p}]=\im\hbar$ imposes $V_t$ to satisfy the condition 
\[
     V_t+\frac{\im\hbar}{2}\Sigma\geq 0, 
\]
that yields the Heisenberg uncertainty relation 
\begin{equation}
\label{heisenberg}
    {\rm det}(V_t)\geq\frac{\hbar^2}{4}~~\Rightarrow~~
     \mean{\Delta\hat{q}_t^2}\mean{\Delta\hat{p}_t^2}\geq\frac{\hbar^2}{4}. 
\end{equation}
This inequality does hold regardless of a measurement setup. 
Hence the following natural question arises. 
Can the Heisenberg limit $\hbar^2/4$ be reached in the linear filtering 
scheme discussed above? 
To answer this important question needs a detailed investigation of 
$V_\infty$, a unique steady solution of the algebraic Riccati equation 
$\dot{V}_\infty=0$ in Eq. (\ref{riccati}). 
(If the Riccati equation does not have such a solution, it implies that 
the estimation fails; we do not take this bad scenario into account.) 
In particular, we aim to get a fundamental lower bound of 
${\rm det}(V_\infty)$ that does not include $C$, $G$, and $\phi$, because 
these terms completely depend on a system under consideration. 
We then obtain the following result. 
\\

{\bf Theorem.} 
Suppose Eq. (\ref{riccati}) has a unique steady solution $V_\infty$. 
Then, the estimation error ${\rm det}(V_\infty)$ has the following 
achievable bounds for any $C, G$, and $\phi$: 
\begin{eqnarray}
& & \hspace*{-1em}
    {\rm det}(V_\infty)\geq 
        \frac{\hbar^2}{4\eta}
           ~~(\mbox{if}~~C_{{\rm r}}^{{\mathsf T}}\Sigma C_{{\rm i}}\leq 0),
\nonumber \\ & & \hspace*{-1em}
    {\rm det}(V_\infty)\geq 
        \frac{\hbar^2}{4}
           ~~(\mbox{if}~~C_{{\rm r}}^{{\mathsf T}}\Sigma C_{{\rm i}}> 0). 
\nonumber
\end{eqnarray}

{\bf Proof.} 
The proof is done by a straightforward calculation. 
Without loss of generality, we can assume that $C_{{\rm r}}$ is normalized: 
$C_{{\rm r}}^{{\mathsf T}}C_{{\rm r}}=1$. 
Let $\bar{C}_{{\rm r}}$ be a unit real vector orthogonal to $C_{{\rm r}}$, i.e., 
$\bar{C}_{{\rm r}}^{{\mathsf T}}\bar{C}_{{\rm r}}=1$ and $C_{{\rm r}}^{{\mathsf T}}\bar{C}_{{\rm r}}=0$, 
and define 
\[
    v_1:=C_{{\rm r}}^{{\mathsf T}}V_\infty C_{{\rm r}},~~
    v_2:=C_{{\rm r}}^{{\mathsf T}}V_\infty\bar{C}_{{\rm r}},~~
    v_3:=\bar{C}_{{\rm r}}^{{\mathsf T}}V_\infty\bar{C}_{{\rm r}}. 
\]
Then, as $(C_{{\rm r}},\bar{C}_{{\rm r}})$ is a $2\times 2$ orthogonal matrix, we have 
\[
    {\rm det}(V_\infty)
    ={\rm det}\Big[\left( \begin{array}{c}
                    C_{{\rm r}}^{{\mathsf T}} \\
                    \bar{C}_{{\rm r}}^{{\mathsf T}} \\
                   \end{array} \right) 
                   V_\infty
                   (C_{{\rm r}},\bar{C}_{{\rm r}})\Big]
    =v_1v_3-v_2^2. 
\]
Furthermore, let us define 
\begin{eqnarray}
& & \hspace*{-1em}
    \left( \begin{array}{cc}
      a_1 & a_2 \\
      a_3 & a_4 \\
    \end{array} \right)
    :=\left( \begin{array}{cc}
       C_{{\rm r}}^{{\mathsf T}}A' C_{{\rm r}} & C_{{\rm r}}^{{\mathsf T}}A' \bar{C}_{{\rm r}} \\
       \bar{C}_{{\rm r}}^{{\mathsf T}}A' C_{{\rm r}} & \bar{C}_{{\rm r}}^{{\mathsf T}}A' \bar{C}_{{\rm r}} \\
      \end{array} \right), 
\nonumber \\ & & \hspace*{-1em}
    \left( \begin{array}{cc}
      d_1 & d_2 \\
      d_2 & d_3 \\
    \end{array} \right)
    :=\left( \begin{array}{cc}
       C_{{\rm r}}^{{\mathsf T}}D C_{{\rm r}} & C_{{\rm r}}^{{\mathsf T}}D \bar{C}_{{\rm r}} \\
       \bar{C}_{{\rm r}}^{{\mathsf T}}D C_{{\rm r}} & \bar{C}_{{\rm r}}^{{\mathsf T}}D \bar{C}_{{\rm r}} \\
      \end{array} \right). 
\nonumber
\end{eqnarray}
Note that $D^{{\mathsf T}}=D$. 
With the above notations, the algebraic Riccati equation $\dot{V}_\infty=0$ 
is reduced to 
\begin{eqnarray}
& & \hspace*{-1em}
\label{ProofRic1}
    2a_1v_1+2a_2v_2+d_1-\frac{4\eta}{\hbar}v_1^2=0, 
\\ & & \hspace*{-1em}
\label{ProofRic2}
    a_3v_1+(a_1+a_4)v_2+a_2v_3+d_2-\frac{4\eta}{\hbar}v_1v_2=0, 
\\ & & \hspace*{-1em}
\label{ProofRic3}
    2a_3v_2+2a_4v_3+d_3-\frac{4\eta}{\hbar}v_2^2=0. 
\end{eqnarray}
Then, adding $v_2^2\times$(\ref{ProofRic1}), 
$-2v_1v_2\times$(\ref{ProofRic2}), and $v_1^2\times$(\ref{ProofRic3}), 
we readily obtain 
\[
    2(v_1v_3-v_2^2)(a_2v_2-a_4v_1)=d_3v_1^2-2d_2v_1v_2+d_1v_2^2. 
\]
This together with Eq. (\ref{ProofRic1}) leads to 
\[
    {\rm det}(V_\infty)
     =\frac{\hbar}{4\eta}\cdot
      \frac{d_3v_1^2-2d_2v_1v_2+d_1v_2^2}
           {v_1^2-\hbar(a_1+a_4)v_1/2\eta-\hbar d_1/4\eta}. 
\]
Note that the denominator is strictly positive from the assumption that the 
Riccati equation has a unique steady solution. 
Now, calculating $d_i$, e.g., 
$d_1=\hbar(1-\eta)(C_{{\rm r}}^{{\mathsf T}}\Sigma C_{{\rm i}})^2$, the numerator of 
${\rm det}(V_\infty)$ is evaluated as 
\begin{eqnarray}
& & \hspace*{-2em}
    d_3v_1^2-2d_2v_1v_2+d_1v_2^2
\nonumber \\ & & \hspace*{-1em}
    =\hbar v_1^2
       +\hbar(1-\eta)\big[(\bar{C}_{{\rm r}}^{{\mathsf T}}\Sigma C_{{\rm i}})v_1
                          -(C_{{\rm r}}^{{\mathsf T}}\Sigma C_{{\rm i}})v_2\big]^2
    \geq \hbar v_1^2, 
\nonumber
\end{eqnarray}
from which we have 
\[
    {\rm det}(V_\infty)\geq
     \frac{\hbar^2}{4\eta}\cdot
      \frac{v_1^2}{v_1^2-\hbar(a_1+a_4)v_1/2\eta-\hbar d_1/4\eta}. 
\]
The right-hand side of the above inequality is further evaluated as follows. 
First, if $a_1+a_4=2(\eta-1)C_{{\rm r}}^{{\mathsf T}}\Sigma C_{{\rm i}}=0$, which implies 
$d_1=0$, we immediately obtain ${\rm det}(V_\infty)\geq\hbar^2/4\eta$. 
Second, if $C_{{\rm r}}^{{\mathsf T}}\Sigma C_{{\rm i}}<0$, which implies $a_1+a_4>0$ and 
$d_1>0$, we have ${\rm det}(V_\infty)>\hbar^2/4\eta$. 
Finally, let us consider the case of $C_{{\rm r}}^{{\mathsf T}}\Sigma C_{{\rm i}}>0$ that 
leads to $a_1+a_4<0$ and $d_1>0$; 
a simple calculation clarifies that the function 
$f(v)=v^2/(v^2+av-b),~(a>0, b>0)$ satisfies $f(v)\geq 4b/(4b+a^2)$ when 
$v>0$ and $v^2+av-b>0$. 
This lower bound becomes $\eta$ in our problem where $a=-\hbar(a_1+a_4)/2\eta$ 
and $b=\hbar d_1/4\eta$. 
As a result, we obtain ${\rm det}(V_\infty)\geq\hbar^2/4$ in this case. 
The achievability of the above lower bounds is discussed in the example 
part. 
$~\blacksquare$
\\

We now give a physical interpretation to the sign of 
$C_{{\rm r}}^{{\mathsf T}}\Sigma C_{{\rm i}}$. 
To do this, let us focus on the matrix $A$, which corresponds to the drift 
term of the quantum dynamics (\ref{linear-qsde}) and the filter 
(\ref{linear-filter}). 
The characteristic polynomial of $A$ is 
$\lambda^2+2(C_{{\rm r}}^{{\mathsf T}}\Sigma C_{{\rm i}})\lambda+
(C_{{\rm r}}^{{\mathsf T}}\Sigma C_{{\rm i}})^2+{\rm det}(G)=0$. 
Hence, $A$ has two stable eigenvalues if and only if the conditions 
\begin{equation}
\label{stability-condition}
    C_{{\rm r}}^{{\mathsf T}}\Sigma C_{{\rm i}}>0,~~~
    (C_{{\rm r}}^{{\mathsf T}}\Sigma C_{{\rm i}})^2+{\rm det}(G)>0
\end{equation}
are satisfied. 
The latter condition is easily attained by making the coefficient of $C$ 
(i.e., the interaction strength) sufficiently large, if the former condition 
is already satisfied. 
Therefore, under the condition $C_{{\rm r}}^{{\mathsf T}}\Sigma C_{{\rm i}}>0$, both the 
quantum dynamics and the filter are (asymptotically) stable in the sense 
that, roughly speaking, those trajectories are constrained around $\hat{x}=0$ 
and $\pi_t(\hat{x})=0$. 
This implies that the fundamental estimation limit $\hbar^2/4\eta$ can be 
violated if the dynamics we aim to track is stable. 
Combining the theorem with the above discussion, we deduce the following 
fact: 
\begin{eqnarray}
& & \hspace*{-1em}
    \mean{\Delta\hat{q}_\infty^2}\mean{\Delta\hat{p}_\infty^2}\geq 
        \frac{\hbar^2}{4\eta}
           ~~(\mbox{if the system is unstable}),
\nonumber \\ & & \hspace*{-1em}
    \mean{\Delta\hat{q}_\infty^2}\mean{\Delta\hat{p}_\infty^2}\geq 
        \frac{\hbar^2}{4}
           ~~(\mbox{if the system is stable}).
\nonumber
\end{eqnarray}

{\bf Remark 1.} 
In practice we cannot construct a perfect measurement apparatus with $\eta=1$. 
Thus, the condition $C_{{\rm r}}^{{\mathsf T}}\Sigma C_{{\rm i}}>0$ is 
clearly preferable from the estimation performance viewpoint. 
Actually, for example when $C_{{\rm r}}^{{\mathsf T}}\Sigma C_{{\rm i}}\leq 0$ 
and $\eta=1/4$, the estimation error is lower bounded by $\hbar^2$, i.e., 
$\mean{\Delta\hat{q}_\infty^2}\mean{\Delta\hat{p}_\infty^2}\geq\hbar^2$, 
which is much bigger than the Heisenberg limit $\hbar^2/4$. 
However, the sign of $C_{{\rm r}}^{{\mathsf T}}\Sigma C_{{\rm i}}$ cannot 
be changed by tuning the Hamiltonian matrix $G$ and the phase-shift $\phi$. 
(Note that 
$C_{{\rm r}}^{{\mathsf T}}\Sigma C_{{\rm i}}
={\rm Re}(C)^{{\mathsf T}}\Sigma{\rm Im}(C)$.) 
In other words, only the interaction term $C$ is the crucial factor that 
determines the estimation limit. 
\\

{\bf Remark 2.} 
The Hamiltonian of the form 
$\hat{H}=\hat{x}^{{\mathsf T}}G\hat{x}/2-\hat{x}^{{\mathsf T}}\Sigma Bu_t$, 
where $B\in{\mathbb R}^2$, allows that the system dynamics 
\[
    d\hat{x}_t=A\hat{x}_tdt+Bu_t
           +\im\Sigma[Cd\hat{B}_t\dgg-C^* d\hat{B}_t] 
\]
can be controlled using a feedback input $u_t\in{\cal Y}_t$. 
For example, the quantum linear quadratic gaussian (LQG) controller 
effectively stabilizes the system. 
However, any control input cannot reduce the estimation limit, because the 
error covariance matrix $V_t$ obeys the same Riccati equation (\ref{riccati}) 
without respect to $B$ and $u_t$. 
\\

We will show that the two bounds in the theorem are tight in a sense that 
there exists at least one example where the equality holds in each case. 
\\

{\bf Example 1.} 
Doherty {\it et. al.} considered in \cite{doherty2} a single particle 
system with the following harmonic oscillator potential and the interaction 
with strength $\alpha>0$: 
\[
    \hat{H}=\frac{m\omega^2}{2}\hat{q}^2+\frac{1}{2m}\hat{p}^2,~~
    \hat{c}=\sqrt{2\alpha}\hat{q}. 
\]
This corresponds to 
\begin{eqnarray}
& & \hspace*{-1em}
    G=\left( \begin{array}{cc}
        m\omega^2 & 0 \\
        0 & 1/m \\
      \end{array} \right),~~
    C=\sqrt{2\alpha}
      \left( \begin{array}{c}
        1 \\
        0 \\
      \end{array} \right), 
\nonumber \\ & & \hspace*{-1em}
    C_{{\rm r}}=\left( \begin{array}{c}
          \sqrt{2\alpha}\cos\phi \\
          0 \\
        \end{array} \right),~~
    C_{{\rm i}}=\left( \begin{array}{c}
          -\sqrt{2\alpha}\sin\phi \\
          0 \\
        \end{array} \right). 
\nonumber
\end{eqnarray}
First, we remark that the Ricatti equation (\ref{riccati}) has a unique 
steady solution $V_\infty$ for all the parameters. 
Then, due to $C_{{\rm r}}^{{\mathsf T}}\Sigma C_{{\rm i}}=0$, the estimation error is 
bounded by 
\[
    {\rm det}(V_\infty)\geq\frac{\hbar^2}{4\eta}~~\Rightarrow~~
     \mean{\Delta\hat{q}_\infty^2}\mean{\Delta\hat{p}_\infty^2}
        \geq\frac{\hbar^2}{4\eta}. 
\]
Actually, the drift matrix $A$ has eigenvalues $\pm\im\omega$, 
implying that the particle is oscillating with frequency $\omega$, and thus 
that the system is not stable. 
Furthermore, in this case, we can obtain a simple explicit form of 
${\rm det}(V_\infty)$: 
\[
    {\rm det}(V_\infty)=\frac{\hbar^2}{4\eta}
       \Big(\frac{1-\eta}{\cos^2\phi}+\eta\Big), 
\]
which attains $\hbar^2/4\eta$ when $\phi=0$. 
Therefore, the lower bound $\hbar^2/4\eta$ is indeed achievable. 
In particular, when $\phi=0$ we have 
\[
    \mean{\Delta\hat{q}_\infty^2}\mean{\Delta\hat{p}_\infty^2}
    =\frac{\hbar^2}{4\eta}+
      \frac{\hbar^2/4\eta}{\sqrt{r_1^2+r_1+\hbar^2/4\eta}}, 
\]
where $r_1=\hbar m\omega^2/8\eta\alpha$. 
Thus, in the limit of $r_1\rightarrow\infty$ the estimation error satisfies 
the minimum uncertainty relation 
$\mean{\Delta\hat{q}_\infty^2}\mean{\Delta\hat{p}_\infty^2}=\hbar^2/4\eta$, 
which further attains the Heisenberg limit 
$\mean{\Delta\hat{q}_\infty^2}\mean{\Delta\hat{p}_\infty^2}=\hbar^2/4$ only 
when $\eta=1$. 
\\

{\bf Example 2.} 
Wiseman and Doherty considered in \cite{wiseman} an atomic system in a 
damped cavity containing an on-threshold parametric down converter that 
realizes 
\[
    \hat{H}=\frac{\beta}{2}(\hat{q}\hat{p}+\hat{p}\hat{q}),~~
    \hat{c}=\gamma(\hat{q}+\im\hat{p}), 
\]
where $\beta>0$ and $\gamma>0$ are parameters. 
We then have 
\begin{eqnarray}
& & \hspace*{-1em}
    G=\left( \begin{array}{cc}
        0 & \beta \\
        \beta & 0 \\
      \end{array} \right),~~
    C=\gamma
      \left( \begin{array}{c}
        1 \\
        \im \\
      \end{array} \right), 
\nonumber \\ & & \hspace*{-1em}
    C_{{\rm r}}=\gamma
        \left( \begin{array}{c}
          \cos\phi \\
          -\sin\phi \\
        \end{array} \right),~~
    C_{{\rm i}}=\gamma
        \left( \begin{array}{c}
          \sin\phi \\
          \cos\phi \\
        \end{array} \right). 
\nonumber
\end{eqnarray}
The Ricatti equation (\ref{riccati}) has a unique steady solution under the 
condition $\beta+\gamma^2>0$, which is already satisfied. 
Then, due to $C_{{\rm r}}^{{\mathsf T}}\Sigma C_{{\rm i}}=\gamma^2>0$, the estimation error 
is lower bounded by $\hbar^2/4$ from the theorem. 
This bound is achievable as in the former example. 
Actually, when $\phi=0$, the off-diagonal term of $V_\infty$ is zero, and 
eventually we have 
\[
    \mean{\Delta\hat{q}_\infty^2}\mean{\Delta\hat{p}_\infty^2}
    =\frac{\hbar^2}{8\eta}\cdot
     \frac{ \sqrt{r_2^2+2(2\eta-1)r_2+1} + r_2+2\eta-1 }
          {1+r_2}, 
\]
where $r_2:=\beta/\gamma^2$. 
Hence, when $\gamma\rightarrow\infty$, which implies that the interaction 
strength is very large, the Heisenberg limit $\hbar^2/4$ is reached.

\end{document}